\RequirePackage{lineno} 
\documentclass[prd,twocolumn,showpacs,preprintnumbers,amsmath,amssymb,superscriptaddress]{revtex4}
\usepackage{graphicx}
\usepackage{amsmath}
\usepackage{amssymb}
\usepackage{afterpage}
\usepackage{graphics}
\usepackage{float}
\usepackage{rotating}
\usepackage{xspace}
\usepackage{dcolumn}
\usepackage{bm} 

\begin{document}

\preprint{FERMILAB-PUB-11-676-E, arXiv:1201.2631 [hep-ex]}

\title{Search for Lorentz invariance and CPT violation with muon antineutrinos in the MINOS Near Detector}         

\newcommand{\Berkeley}{Lawrence Berkeley National Laboratory, Berkeley, California, 94720 USA}
\newcommand{\Cambridge}{Cavendish Laboratory, University of Cambridge, Madingley Road, Cambridge CB3 0HE, United Kingdom}
\newcommand{\FNAL}{Fermi National Accelerator Laboratory, Batavia, Illinois 60510, USA}
\newcommand{\RAL}{Rutherford Appleton Laboratory, Science and Technologies Facilities Council, OX11 0QX, United Kingdom}
\newcommand{\UCL}{Department of Physics and Astronomy, University College London, Gower Street, London WC1E 6BT, United Kingdom}
\newcommand{\Caltech}{Lauritsen Laboratory, California Institute of Technology, Pasadena, California 91125, USA}
\newcommand{\Alabama}{Department of Physics and Astronomy, University of Alabama, Tuscaloosa, Alabama 35487, USA}
\newcommand{\ANL}{Argonne National Laboratory, Argonne, Illinois 60439, USA}
\newcommand{\Athens}{Department of Physics, University of Athens, GR-15771 Athens, Greece}
\newcommand{\NTUAthens}{Department of Physics, National Tech. University of Athens, GR-15780 Athens, Greece}
\newcommand{\Benedictine}{Physics Department, Benedictine University, Lisle, Illinois 60532, USA}
\newcommand{\BNL}{Brookhaven National Laboratory, Upton, New York 11973, USA}
\newcommand{\CdF}{APC -- Universit\'{e} Paris 7 Denis Diderot, 10, rue Alice Domon et L\'{e}onie Duquet, F-75205 Paris Cedex 13, France}
\newcommand{\Cleveland}{Cleveland Clinic, Cleveland, Ohio 44195, USA}
\newcommand{\Delhi}{Department of Physics \& Astrophysics, University of Delhi, Delhi 110007, India}
\newcommand{\GEHealth}{GE Healthcare, Florence South Carolina 29501, USA}
\newcommand{\Harvard}{Department of Physics, Harvard University, Cambridge, Massachusetts 02138, USA}
\newcommand{\HolyCross}{Holy Cross College, Notre Dame, Indiana 46556, USA}
\newcommand{\Houston}{Department of Physics, University of Houston, Houston, Texas 77204, USA}
\newcommand{\IIT}{Department of Physics, Illinois Institute of Technology, Chicago, Illinois 60616, USA}
\newcommand{\Iowa}{Department of Physics and Astronomy, Iowa State University, Ames, Iowa 50011 USA}
\newcommand{\Indiana}{Indiana University, Bloomington, Indiana 47405, USA}
\newcommand{\ITEP}{High Energy Experimental Physics Department, ITEP, B. Cheremushkinskaya, 25, 117218 Moscow, Russia}
\newcommand{\JMU}{Physics Department, James Madison University, Harrisonburg, Virginia 22807, USA}
\newcommand{\LASL}{Nuclear Nonproliferation Division, Threat Reduction Directorate, Los Alamos National Laboratory, Los Alamos, New Mexico 87545, USA}
\newcommand{\Lebedev}{Nuclear Physics Department, Lebedev Physical Institute, Leninsky Prospect 53, 119991 Moscow, Russia}
\newcommand{\LLL}{Lawrence Livermore National Laboratory, Livermore, California 94550, USA}
\newcommand{\LosAlamos}{Los Alamos National Laboratory, Los Alamos, New Mexico 87545, USA}
\newcommand{\MIT}{Lincoln Laboratory, Massachusetts Institute of Technology, Lexington, Massachusetts 02420, USA}
\newcommand{\Minnesota}{University of Minnesota, Minneapolis, Minnesota 55455, USA}
\newcommand{\Crookston}{Math, Science and Technology Department, University of Minnesota -- Crookston, Crookston, Minnesota 56716, USA}
\newcommand{\Duluth}{Department of Physics, University of Minnesota -- Duluth, Duluth, Minnesota 55812, USA}
\newcommand{\Ohio}{Center for Cosmology and Astro Particle Physics, Ohio State University, Columbus, Ohio 43210 USA}
\newcommand{\Otterbein}{Otterbein College, Westerville, Ohio 43081, USA}
\newcommand{\Oxford}{Subdepartment of Particle Physics, University of Oxford, Oxford OX1 3RH, United Kingdom}
\newcommand{\PennState}{Department of Physics, Pennsylvania State University, State College, Pennsylvania 16802, USA}
\newcommand{\PennU}{Department of Physics and Astronomy, University of Pennsylvania, Philadelphia, Pennsylvania 19104, USA}
\newcommand{\Pittsburgh}{Department of Physics and Astronomy, University of Pittsburgh, Pittsburgh, Pennsylvania 15260, USA}
\newcommand{\IHEP}{Institute for High Energy Physics, Protvino, Moscow Region RU-140284, Russia}
\newcommand{\Rochester}{Department of Physics and Astronomy, University of Rochester, New York 14627 USA}
\newcommand{\RoyalH}{Physics Department, Royal Holloway, University of London, Egham, Surrey, TW20 0EX, United Kingdom}
\newcommand{\Carolina}{Department of Physics and Astronomy, University of South Carolina, Columbia, South Carolina 29208, USA}
\newcommand{\SLAC}{Stanford Linear Accelerator Center, Stanford, California 94309, USA}
\newcommand{\Stanford}{Department of Physics, Stanford University, Stanford, California 94305, USA}
\newcommand{\StJohnFisher}{Physics Department, St. John Fisher College, Rochester, New York 14618 USA}
\newcommand{\Sussex}{Department of Physics and Astronomy, University of Sussex, Falmer, Brighton BN1 9QH, United Kingdom}
\newcommand{\TexasAM}{Physics Department, Texas A\&M University, College Station, Texas 77843, USA}
\newcommand{\Texas}{Department of Physics, University of Texas at Austin, 1 University Station C1600, Austin, Texas 78712, USA}
\newcommand{\TechX}{Tech-X Corporation, Boulder, Colorado 80303, USA}
\newcommand{\Tufts}{Physics Department, Tufts University, Medford, Massachusetts 02155, USA}
\newcommand{\UNICAMP}{Universidade Estadual de Campinas, IFGW-UNICAMP, CP 6165, 13083-970, Campinas, SP, Brazil}
\newcommand{\UFG}{Instituto de F\'{i}sica, Universidade Federal de Goi\'{a}s, CP 131, 74001-970, Goi\^{a}nia, GO, Brazil}
\newcommand{\USP}{Instituto de F\'{i}sica, Universidade de S\~{a}o Paulo,  CP 66318, 05315-970, S\~{a}o Paulo, SP, Brazil}
\newcommand{\Warsaw}{Department of Physics, University of Warsaw, Ho\.{z}a 69, PL-00-681 Warsaw, Poland}
\newcommand{\Washington}{Physics Department, Western Washington University, Bellingham, Washington 98225, USA}
\newcommand{\WandM}{Department of Physics, College of William \& Mary, Williamsburg, Virginia 23187, USA}
\newcommand{\Wisconsin}{Physics Department, University of Wisconsin, Madison, Wisconsin 53706, USA}
\newcommand{\deceased}{Deceased.}

\affiliation{\ANL}
\affiliation{\Athens}
\affiliation{\BNL}
\affiliation{\Caltech}
\affiliation{\Cambridge}
\affiliation{\UNICAMP}
\affiliation{\FNAL}
\affiliation{\UFG}
\affiliation{\Harvard}
\affiliation{\HolyCross}
\affiliation{\IIT}
\affiliation{\Indiana}
\affiliation{\Iowa}
\affiliation{\UCL}
\affiliation{\Minnesota}
\affiliation{\Duluth}
\affiliation{\Otterbein}
\affiliation{\Oxford}
\affiliation{\Pittsburgh}
\affiliation{\RAL}
\affiliation{\USP}
\affiliation{\Carolina}
\affiliation{\Stanford}
\affiliation{\Sussex}
\affiliation{\TexasAM}
\affiliation{\Texas}
\affiliation{\Tufts}
\affiliation{\Warsaw}
\affiliation{\WandM}

\author{P.~Adamson}
\affiliation{\FNAL}






\author{D.~S.~Ayres}
\affiliation{\ANL}





\author{G.~Barr}
\affiliation{\Oxford}









\author{M.~Bishai}
\affiliation{\BNL}

\author{A.~Blake}
\affiliation{\Cambridge}


\author{G.~J.~Bock}
\affiliation{\FNAL}

\author{D.~J.~Boehnlein}
\affiliation{\FNAL}

\author{D.~Bogert}
\affiliation{\FNAL}




\author{S.~V.~Cao}
\affiliation{\Texas}

\author{S.~Cavanaugh}
\affiliation{\Harvard}



% ...

\collaboration{The MINOS Collaboration}
\noaffiliation
\date{\today}          

\begin{abstract}
We have searched for sidereal variations in the rate of antineutrino interactions in the MINOS Near Detector. Using antineutrinos  produced by the NuMI beam, we find no statistically significant sidereal modulation in the rate.  When this result is placed in the context of the Standard Model Extension theory we are able to place upper limits on the coefficients defining the theory. These limits are used in combination with the results from an earlier analysis of MINOS neutrino data to further constrain the coefficients.
\end{abstract}

\maketitle

Central to both the Standard Model (SM) and General Relativity are the principles of Lorentz and CPT invariance.  The Standard Model Extension (SME)~\cite{CKcomb, others} provides a framework for potential Lorentz invariance violation (LV) and CPT invariance violation (CPTV) in the SM and suggests such violations could occur at the Planck scale, $10^{19}$ GeV. These violations could manifest themselves at observable energies through several unconventional phenomena.  One possibility is a potential dependence of the neutrino and antineutrino oscillation probability on the direction of propagation with respect to the Sun-centered inertial frame in which the SME is formulated~\cite{KM}.  An experiment that has both its antineutrino beam and detector fixed on the Earth's surface could then observe a sidereal variation in the number of antineutrinos detected from the beam.  

MINOS is such an experiment~\cite{Michael:2008bc}.  It uses Fermilab's NuMI neutrino beam~\cite{numi} and two detectors. The MINOS Near Detector (ND) is located 1.04~km from the neutrino production target and the Far Detector (FD) is located 735~km from the production target. The NuMI beam can be configured to enhance the muon antineutrino component for high statistics studies using antineutrinos.  Both detectors are magnetized to approximately 1.4~T, allowing for the discrimination of $\mu^{+}$ produced in charged-current (CC) antineutrino interactions from $\mu^{-}$ produced in CC neutrino interactions.  Because of their different baselines, the ND and FD are sensitive to different limits of the general SME formulated for the neutrino sector.  The predicted SME effects for baselines of about 1~km are independent of neutrino mass~\cite{KM2}, while for long baselines the effects are a perturbation on the standard mass oscillation scenario~\cite{dkm}.  MINOS has found no statistically significant evidence for these effects with neutrinos observed in either its ND~\cite{ndlv} or FD~\cite{paper2}.  The high data rate in the ND allows us to expand our search to include antineutrinos produced by the NuMI beam.

According to the SME, for short baselines the probability that a $\bar \nu_\mu$ oscillates to flavor $\bar \nu_x$, where $x$ is $e$ or $\tau$, over a distance $L$ from its production to its detection due to LV and CPTV is given by~\cite{KM} 
\begin{eqnarray}
\label{eq:osc}
P_{{\bar \nu}_{\mu} \rightarrow {\bar \nu_x}} &\simeq& L^2 [( {\mathcal C})_{\bar x \bar \mu} +  
({\mathcal A_c})_{\bar x \bar \mu} \cos{(\omega_\oplus T_\oplus)} \nonumber \\ 
& & + 
({\mathcal A_s})_{\bar x \bar \mu}   \sin{(\omega_\oplus T_\oplus)} + 
({\mathcal B_c})_{\bar x \bar \mu}   \cos{(2 \omega_\oplus T_\oplus)}  \nonumber \\
& & + ({\mathcal B_s})_{\bar x \bar \mu}  \sin{(2 \omega_\oplus T_\oplus)}]^2,
\end{eqnarray}
where $\omega_\oplus= 2 \pi/(23^h 56^m 04.0982^s)$ is the Earth's sidereal frequency, and $T_\oplus$ is the local sidereal time of the antineutrino event.  The average value of $L$ is $750$~m for antineutrinos that are produced by hadron decays in the NuMI beam and that interact in the ND. The magnitudes of the parameters in Eq.~(\ref{eq:osc}) depend on the neutrino energy, the SME coefficients described below and the direction of the neutrino propagation in the coordinate system fixed on the rotating Earth.   The direction vectors are defined by the colatitude of the NuMI beam line $\chi = (90^\circ -$ latitude) = 42.17973347$^{\circ}$, the beam zenith angle $\theta=93.2745^\circ$ defined from the $z$-axis which points up toward the local zenith, and the beam azimuthal angle $\phi=203.909^\circ$ measured counterclockwise from the $x$-axis chosen to lie along the detector's long axis.  

Equation~(\ref{eq:osc}) for antineutrinos in the ND is identical to the oscillation probability equation for neutrinos in the ND~\cite{ndlv}, with the parameters  $({\mathcal{A}_{c}})_{\bar x  \bar \mu},\dots,({\mathcal{B}_{s}})_{\bar x \bar \mu}$ replacing their counterparts $({\mathcal{A}_{c}})_{x  \mu}, \dots, ({\mathcal{B}_{s}})_{ x  \mu}$.  The parameter $({\mathcal{C}})_{\bar x  \bar \mu}$ similarly replaces $({\mathcal{C}})_{ x  \mu}$, but does not play a role in the sidereal analysis and is not considered further.

In the SME theory the antineutrino oscillation parameters $({\mathcal{A}_{c}})_{\bar x  \bar \mu}, \dots, ({\mathcal{B}_{s}})_{\bar x \bar \mu}$ are functions of the coefficients $(a_L)^\alpha_{ab}$ and $(c_L)^{\alpha \beta}_{ab}$~\cite{KM}. There are 36 of these coefficients: the real and imaginary components of $(a_L)^{X}$, $(a_L)^{Y}$, $(c_L)^{TX}$, $(c_L)^{TY}$, $(c_L)^{XX}$, $(c_L)^{YY}$, $(c_L)^{XY}$, $(c_L)^{YZ}$, $(c_L)^{XZ}$ for $\bar\nu_\mu \rightarrow \bar\nu_e$ and $\bar\nu_\mu \rightarrow \bar\nu_\tau$.  Further, these same 36 coefficients also describe the neutrino oscillation parameters $({\mathcal{A}_{c}})_{x  \mu}, \dots, ({\mathcal{B}_{s}})_{ x  \mu}$.  However, the way in which the real and imaginary components of the $(a_L)^{\alpha}_{ab}$ and $(c_L)_{ab}^{\alpha\beta}$ coefficients participate in the $({\mathcal{A}_{c}})_{\bar x  \bar \mu}, \dots, ({\mathcal{B}_{s}})_{\bar x \bar \mu}$ parameters is different from the way in which they participate in $({\mathcal{A}_{c}})_{x  \mu}, \dots, ({\mathcal{B}_{s}})_{ x  \mu}$.  The reason for the difference is the decomposition of the $(a_L)^{\alpha}$ and $(c_L)^{\alpha\beta}$ coefficients into real and imaginary components.  For neutrinos 
\begin{eqnarray}
(a_L)^{\alpha}_{ab}      &=& {\mathcal Re} (a_L)^{\alpha}_{ab} + i \,{\mathcal Im}( a_L)^{\alpha}_{ab}  \nonumber \\
(c_L)^{\alpha\beta}_{ab} &=& {\mathcal Re} (c_L)^{\alpha\beta}_{ab} + i \,{\mathcal Im}( c_L)^{\alpha\beta}_{ab},
\label{eq:neuts}
\end{eqnarray}
and for antineutrinos
\begin{eqnarray}
(a_R)^{\alpha}_{\bar a \bar b}      &=& -{\mathcal Re} (a_L)^{\alpha}_{ab} + i \,{\mathcal Im}( a_L)^{\alpha}_{ab}  \nonumber \\
(c_R)^{\alpha\beta}_{\bar a \bar b} &=& {\mathcal Re} (c_L)^{\alpha\beta}_{ab} - i \,{\mathcal Im}( c_L)^{\alpha\beta}_{ab}.
\label{eq:antineuts}
\end{eqnarray}
The subscript ``$L$'' in Eq.~(\ref{eq:neuts}) reflects the left-handed nature of neutrinos while the subscript ``$R$'' in Eq.~(\ref{eq:antineuts}) reflects the right-handed nature of antineutrinos. There is a possibility that fortuitous cancellations in the many SME coefficients describing neutrino oscillations could have masked the sidereal signal for which we were searching.  However, the different dependencies of the parameters for neutrinos and antineutrinos on the SME coefficients suggest that  it is unlikely that a second set of fortuitous cancellations would also mask an LV sidereal signal for antineutrinos.  

Our primary motivation for this analysis is to explore a new window into LV with antineutrinos.  Furthermore this analysis sheds light on whether cancellations among the SME coefficients can affect the results.  If MINOS is sensitive to sidereal effects resulting from LV in the neutrino sector and these effects are being masked by accidental cancellations, then this antineutrino analysis would find them.  On the other hand, if we find no significant evidence for a sidereal signal in antineutrinos, we can use our results to improve the MINOS upper limits on the SME coefficients we previously found with neutrinos since the same coefficients describe both neutrino and antineutrino oscillations.

We applied standard MINOS beam and data quality selection~\cite{Adamson:2007gu} to select beam spills for the analysis.  We also applied data quality cuts to remove data where there were cooling system problems, magnetic coil problems, or an incorrectly configured readout trigger.  

Two independent periods of muon antineutrino data taking are combined to comprise the data set for this analysis.  Table~\ref{table:runParam} gives the run dates, number of protons incident on the target (POT), and the number of CC events remaining in the sample after all selections have been made, N$_{CC}$.  The events were selected following the prescription of a previous MINOS analysis~\cite{Adamson:2011fa}.  Studies have shown that the mean number of antineutrinos per POT in the ND has remained stable to about 1\% throughout the data taking. Our previous analysis used $3.54 \times 10^6$ muon neutrinos observed in the ND~\cite{ndlv}.

\begin{table}[h]
\caption{\label{table:runParam} Antineutrino Data Sample.}
\begin{tabular}{|c|c|c|}
\hline \hline
  ~~~~Run Dates~~~~& ~POT~ & N$_{CC}$~\\  \hline \hline

Sep09 -- Mar10 & $~~1.67 \times 10^{20}$  ~~& ~~637,805~~  \\

Nov10 -- Jan11 & ~~$0.98 \times 10^{20}$  ~~& ~~379,877~~  \\

\hline 

Total   & ~~ $2.65 \times 10^{20}$ ~~ & ~~ 1,017,682~~  \\

\hline \hline

\end{tabular}

\end{table}

We used the ratio of the events observed to the number of POT recorded as a function of sidereal time as the normalized quantity in which to search for sidereal variations.  We implemented the search for a sidereal signal as a blind analysis where we only examined the event rate for the data once the analysis procedures were determined.  We used the sidereal time distribution of the beam spills and the total number of antineutrino events in the data set as inputs to generate $10^{4}$ numerical experiments that simulated the data set without a sidereal signal.  We then performed a Fourier analysis on these simulated experiments to establish the search criteria needed to find a sidereal signal. 

We constructed our simulated experiments based on the local sidereal time (LST) distribution, $T_{\oplus}$, of the beam spills converted to local sidereal phase (LSP), where $\text{LSP}=\mod(T_{\oplus} \omega_\oplus/2\pi)$.  To generate this histogram, we converted the time of the extraction magnet signal that initiates each spill, as recorded by a GPS unit, into LST in standard ways~\cite{pawyc}.  The GPS time is accurate to 200~ns~\cite{minosToF} and event times were not corrected for the time within the 10~$\mu$s spill.  We then computed the LSP for each beam spill and entered it into a histogram with 32 bins ranging from $0-1$ in LSP.  We chose this binning because the Fast Fourier Transform (FFT) algorithm used to look for sidereal variations works most efficiently for $2^{\mathcal N}$ bins~\cite{numrec}.   Since Eq.~(\ref{eq:osc}) only puts power into the four harmonic terms $\omega_{\oplus}T_{\oplus}, \ldots, 4\omega_{\oplus}T_{\oplus}$, we adopted ${\mathcal N} = 5$ as the binning that retains these harmonic terms while still providing sufficient resolution in sidereal time to detect a signal.   Each phase bin spans 0.031 in LSP or 45 minutes in sidereal time.  

To construct the simulated experiments we took each spill in the data set one at a time and randomly assigned a new LSP for the spill from the LSP distribution of all spills.  We assigned the number of POT in the spill to one histogram in LSP using the newly assigned phase and then checked whether any antineutrino events were recorded for the spill.  If so, we put those events in a second histogram using the same LSP.  This procedure ensured that the correlation between POT and events observed in the ND for each spill was retained.  By the end of the simulation, we have two histograms: one with POT as a function of LSP and one with the events as a function of LSP.  By picking spill times out of the LSP distribution for the data, we are assured that both histograms have their entries distributed properly in LSP.  In addition, we guaranteed that no sidereal signal is present in the simulated experiments since any correlation between the data spills is removed.  We took the ratio of these two histograms to obtain the rate histogram for the simulated experiment. 

\begin{figure}[h]
\centerline{\includegraphics[width=3.25in]{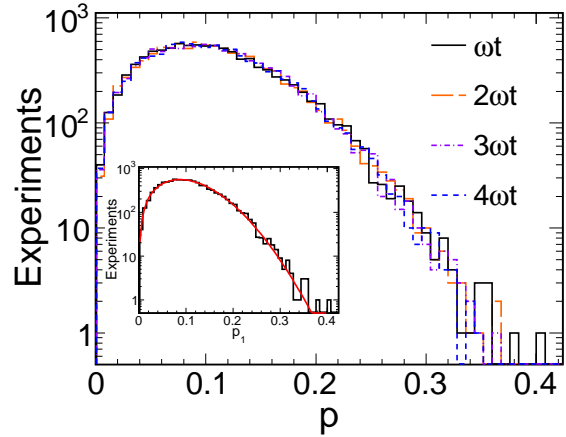}}
\caption{\label{fig:mc_power}  The distributions for the quadratic sum of powers  $p_1, \ldots, p_4$ from the FFT analysis of $10^{4}$ simulated experiments without a sidereal signal. The inset shows the distribution for $p_{1}$ with a fit to a Rayleigh distribution having $\sigma = 0.09$ superposed.}
\end{figure}

We next performed an FFT on each simulated rate histogram and computed the power in the four harmonic terms ($\omega_{\oplus}T_{\oplus}, \ldots, 4\omega_{\oplus}T_{\oplus}$) appearing in the oscillation probability, Eq.(\ref{eq:osc}).   Let $S_1$ be the power returned by the FFT for the first harmonic term  $\sin{(\omega_{\oplus}T_{\oplus}})$ and  $C_1$ be the power returned for the first harmonic term  $\cos{(\omega_{\oplus}T_{\oplus}})$; similarly define ($S_2$,  $C_2$), $\dots$,( $S_4$,  $C_4$). Then the statistics we used in our search are
\begin{equation}
p_1 = \sqrt{S_1^2 + C_1^2}, \ldots, p_4 = \sqrt{S_4^2 + C_4^2}.
\end{equation}
We added the powers in quadrature to eliminate the effect of the arbitrary choice of a zero point in phase at $0^h$ LST.  Fig.~\ref{fig:mc_power} shows the distribution of $p_1, \ldots, p_4$ for the $10^{4}$ simulated experiments.
The distributions for  $p_1, \ldots, p_4$ are quite similar.  These distributions are well described by a Rayleigh distribution with $\sigma = 0.09$, showing that the powers for the sine and cosine terms of the various harmonics are uncorrelated and normally distributed in the experiments.  

Our threshold for signal detection in any harmonic is the quadratic power $p(\text{FFT})$ that is greater than 99.7\% of the entries in its $p_1, \ldots, p_4$ histogram.  We take these signal detection thresholds as the 99.7\% confidence level (C.L.) for the probability that a measured quadratic sum of powers for any harmonic was {\it not} drawn from a distribution having a sidereal signal. These thresholds are 0.30, 0.30, 0.29, and 0.31 for $p_1, \ldots, p_4$, respectively and we adopt $p_{th} = 0.31$ as the overall detection threshold.

We determined the minimum detectable sidereal modulation for this analysis by injecting a sidereal signal of the form $A\sin(\omega_{\oplus} T)$, where $A$ is a fraction of the mean event rate, into a new set of $10^{4}$ simulated experiments and repeating the FFT analysis.  We found that every experiment gave $p_{1} \geq 0.31$ when $A = 0.8\%$ of the mean rate.  Thus, this analysis is sensitive to sub-percent level sidereal variations in the mean event rate.

\begin{figure}[h]
\centerline{\includegraphics[width=3.25in]{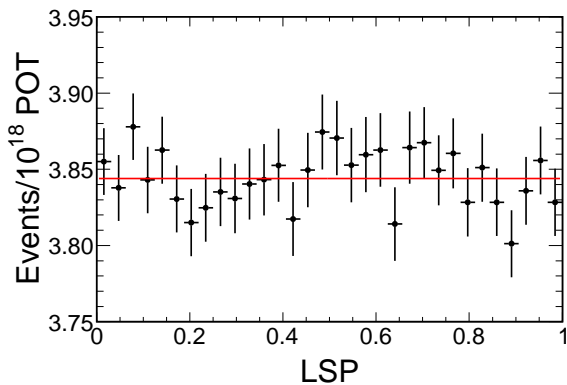}}
\caption{\label{fig:data_rate}  The phase diagram of the CC antineutrino event rate for the ND data.  The mean rate of 3.84 events per $10^{15}$~POT is superposed and has $\chi^{2}/ndf = 21.7/31$.}
\end{figure}

Once the threshold for the signal detection was determined we performed the FFT analysis using the actual data event rate as a function of LSP shown in  Fig.~\ref{fig:data_rate}. 
The results for  $p_1, \dots, p_4$  are given in Table~\ref{table:dataPower}.  This table also shows the probability, $\cal{P}_{F}$, that the measured power is due to a noise fluctuation.  $\cal{P}_F$ is the probability of drawing a value of $p_1, \dots,  p_4$ from the parent distribution in Fig.~\ref{fig:mc_power} at least as large as found in the data.
\begin{table}[h]
\caption{\label{table:dataPower} Results for the $p_1, \dots, p_4$ statistics for the data shown in Fig.~\ref{fig:data_rate}.  The third column gives the probability, $\cal{P}_F$, that the measured power is due to a noise fluctuation. }
\begin{tabular}{|c|c|c|}
\hline\hline
~~Statistic~~ & ~~$p(\text{FFT})$ ~~& ~~~~~$\cal{P}_F$~~~~~ \\ 
\hline
$p_1$ &  0.12 & 0.42\\ 
$p_2$  & 0.17 & 0.16 \\
$p_3$  & 0.13 & 0.35 \\
$p_4$  & 0.10 & 0.54 \\
\hline \hline
\end{tabular}
\end{table}
As none of the values $p_1, \dots, p_4$ exceed our detection threshold, we find no evidence for a sidereal signal in the antineutrino data set.  

 We investigated the sensitivity of our results to several sources of systematic uncertainties.  In the previous MINOS analyses~\cite{ndlv, paper2}, the NuMI target was observed to have degraded, causing a drop in the event rate throughout the exposure.  Because of this degradation, we examined how linear changes in the event rate over time would affect the determination of the detection thresholds and found such changes had no effect.  The NuMI target was replaced between the data taking period of the previous analyses and this analysis.  The new target did not show evidence of degradation during the course of its exposure.   Given that systematic changes in the event rate were shown not to affect the previous results and that there is no evidence for such changes in these data, this source of systematic uncertainty is negligible.  
 
Potential differences in the event rate for data taken during the solar day compared to the solar night are another possible source of systematic uncertainty.  We looked for these effects by searching for systematic differences in the event rate as a function of solar diurnal phase.  These rates are consistent with no diurnal variations and we conclude that diurnal effects are not masking a true sidereal signal in the data.  

There is a known $\pm 1$\% uncertainty in the recorded number of POT per spill~\cite{Adamson:2007gu} that could introduce a modulation that would mask a sidereal signal.  We introduced random variations of this scale in the number of POT recorded from each spill and repeated the FFT analysis.  We observed no change in the detection threshold due to these variations.  Moreover, we observed no changes in the detection threshold due to long term drifts of the size $\pm 5$\% over six months.  Thus we conclude that the POT counting uncertainties cannot mask a sidereal signal.  

As first pointed out by Compton and Getting~\cite{CG}, atmospheric effects can mimic a sidereal modulation if there were a solar diurnal modulation in the event rate that beats with a yearly modulation.  Following the methods described in~\cite{slmCG}, we found the amplitude of the potential faux sidereal modulation would be only $0.5\%$ of our minimum detectable modulation and therefore would not mask a sidereal signal that MINOS could detect.

In the absence of a sidereal signal, we can determine the  99.7\% C.L. upper limits on the SME coefficients ${\mathcal Re}(a_L)^{\alpha}_{ab}$, ${\mathcal Im}(a_L)^{\alpha}_{ab}$,  ${\mathcal Re}(c_L)^{\alpha\beta}_{ab}$, and  ${\mathcal Im}(c_L)^{\alpha\beta}_{ab}$ using the MINOS Monte Carlo simulation~\cite{Adamson:2007gu}.  In this simulation, events are generated by modeling the NuMI beam line, including hadron production by the 120 GeV$/c$ protons, propagation of the hadrons through the focusing elements and 675 m decay pipe to the beam absorber, and the calculation of the probability that any neutrinos generated traverse the ND.  The ND neutrino event simulation takes the neutrinos from the NuMI simulation, along with weights determined by decay kinematics, and uses this information as input into the simulation of the ND.  

We determined the confidence limit for an SME coefficient by simulating a set of experiments in which we set all but this one coefficient to zero.  For the first simulated experiment, we injected a negligible LV signal into the simulation and constructed the resulting LSP histogram.  We calculated the survival probability for each antineutrino from its energy, the distance it travels to the ND in the simulation and a value for the magnitude of the SME coefficient causing a negligible LV signal.   We used this simulated LSP histogram to compute $p_1, \ldots, p_4$ for the experiment.   We repeated the simulation 1000 times to obtain the average value of each $p_1, \ldots, p_4$ statistic for the value of the chosen SME coefficient.  We then increased the value of the SME coefficient and recomputed the average value of each $p_1, \ldots, p_4$ for a second set of experiments.  We continued the process of increasing the value of the SME coefficient until the largest average value of any $p_1, \ldots, p_4$ crossed the detection threshold of 0.31.  We took this value of the SME coefficient to be its 99.7\% C.L.  upper limit. We then computed upper limits for the remaining SME coefficients in the same way.

The 99.7\% C.L. upper limit of the SME coefficients are given in Table~\ref{table:limits1}.  These limits were cross-checked by simulating 1000 experiments for each coefficient in the table, where that coefficient was set to the determined limit and the rest were set to zero.  The distributions of the $p_{1}, \ldots, p_{4}$ statistics for these experiments showed the measured values in Table~\ref{table:dataPower} were excluded at more than the 99.7\% C.L.
\begin{table}[h]
\caption{\label{table:limits1} The 99.7\% C.L.  upper limit on SME Coefficients for $\nu_{\bar \mu} \rightarrow \nu_{\bar x}$; $(a_L)^{\alpha}$ have units [GeV] and $(c_L)^{\alpha\beta}$ are unitless.} 
\begin{tabular}{|lc|lc|} 

\hline \hline


$(a_L)^{X}$ &  $3.3 \times 10^{-20}$ & $(a_L)^Y$ & $3.3 \times 10^{-20}$ \\  


$(c_L)^{TX}$ &  $1.5 \times 10^{-21}$ & $(c_L)^{TY}$ &  $1.5 \times 10^{-21}$ \\

$(c_L)^{XX}$ &  $7.8 \times 10^{-21}$ & $(c_L)^{YY}$ &  $7.8 \times 10^{-21}$ \\

$(c_L)^{XY}$ & $3.9 \times 10^{-21}$ & $(c_L)^{YZ}$ &  $2.3\times 10^{-21}$ \\

$(c_L)^{XZ}$ &  $2.3 \times 10^{-21}$ & ~~--~~ &~~~ -- ~~~\\

\hline \hline

\end{tabular}

\end{table}
This table has the same form as the 99.7\% C.L. tables in \cite{ndlv,paper2}.  We point out that for this analysis, as for the previous ND neutrino analysis~\cite{ndlv}, each limit in this table actually represents  the 99.7\% C.L. upper limit on 4 SME coefficients.  For $(a_L)^X$ these are: ${\mathcal Re}(a_L)^{X}_{e \mu}$, ${\mathcal Im}(a_L)^{X}_{ e \mu}$, ${\mathcal Re}(a_L)^{X}_{\mu \tau }$, and ${\mathcal Im}(a_L)^{X}_{ \mu \tau}$.  Similarly, $(a_L)^{Y}$, $(c_L)^{TX}, \ldots, (c_L)^{XZ}$ represent limits on 4 SME coefficients (the ${\mathcal Re}$ and ${\mathcal Im}$ parts of the coefficients for $\bar\nu_\mu \rightarrow \bar\nu_e$ and $\bar\nu_\mu \rightarrow \bar\nu_\tau$).  
 
By setting all but one SME coefficient to zero to determine its confidence limit, our method is based on the premise that our null detection does not result from fortuitous cancellations of SME coefficients that hide a signal of oscillation terms in Eq.~(\ref{eq:osc}).  Since the number of SME coefficients is large, this could be an issue.  In fact, we raised this issue in \cite{ndlv,paper2} when we determined confidence limits based on our null detections with neutrinos. But when taken together, the null searches for a sidereal signal with both neutrinos and antineutrinos make it clear that fortuitous cancellations are quite unlikely.  Although both neutrino and antineutrino oscillations are described by the same SME coefficients, the oscillation parameters for neutrinos and antineutrinos have different, nonlinear dependencies on them.  Both sets of oscillation parameters would independently have to cancel.  We conclude that our method for determining the limits is sound.

For the 9 SME coefficients ${\mathcal Re}(a_L)^{\alpha}$ and ${\mathcal Re}(c_L)^{\alpha\beta}$ for the channel $\nu_\mu \rightarrow \nu_\tau$, the limits found in \cite{paper2} are the most sensitive we can determine with our analyses of the MINOS neutrino and antineutrino data.  For the remaining 27 SME coefficients, however, we can improve the limits by combining the results from \cite{ndlv} with those in Table~\ref{table:limits1}.  Let $(CL)_\nu$ be the 99.7\% C.L. upper limit on an SME coefficient determined in \cite{ndlv} and $(CL)_{\bar \nu}$ the 99.7\% C.L. upper limit determined here.  We combine the two limits as 
\begin{displaymath}
1/(CL)^2 = 1/(CL)_\nu^2 + 1/(CL)_{\bar \nu}^2,
\end{displaymath}
where $(CL)$ is the combined 99.7\% C.L. upper limit. The most sensitive upper limits we have determined with the MINOS neutrino and antineutrino data are given in Table~\ref{table:limits2}.
\begin{table}[h]
\caption{\label{table:limits2} The most sensitive 99.7\% C.L. upper limits on the SME coefficients determined by MINOS neutrino and antineutrino data; $(a_L)^{\alpha}_{ab}$ have units [GeV] and $(c_L)^{\alpha\beta}_{ab}$ are unitless. Unless otherwise indicated, the limits were determined using ND data.} 

\begin{tabular}{| l | lc|lc |} 
\hline \hline

Coefficient &~~$ab$ &limit~~~~~~&~~$ab$&limit~~~~~~\\

\hline \hline

${\mathcal Re}(a_{L})^{X}_{ab}$~~~&~~$e\mu$~~~~&$2.2\times10^{-20}$~~~~&~~$\mu\tau\footnotemark[6]$~~~~&$5.9\times10^{-23}$ \\
${\mathcal Im}(a_{L})^{X}_{ab}$~~~&~~$e\mu$~~~~&$2.2\times10^{-20}$~~~~&~~$\mu\tau$&$2.2\times10^{-20}$ \\ \hline
${\mathcal Re}(a_{L})^{Y}_{ab}$~~~&~~$e\mu$~~~~&$2.2\times10^{-20}$~~~~&~~$\mu\tau\footnotemark[6]$~~~~&$6.1\times10^{-23}$ \\
${\mathcal Im}(a_{L})^{Y}_{ab}$~~~&~~$e\mu$~~~~&$2.2\times10^{-20}$~~~~&~~$\mu\tau$~~~~&$2.2\times10^{-20}$ \\ \hline
${\mathcal Re}(c_{L})^{TX}_{ab}$~~~&~~$e\mu$~~~~&$9.0\times10^{-23}$~~~~&~~$\mu\tau\footnotemark[6]$~~~~&$0.5\times10^{-23}$ \\
${\mathcal Im}(c_{L})^{TX}_{ab}$~~~&~~$e\mu$~~~~&$9.0\times10^{-23}$~~~~&~~$\mu\tau$~~~~&$9.0\times10^{-23}$ \\ \hline
${\mathcal Re}(c_{L})^{TY}_{ab}$~~~&~~$e\mu$~~~~&$9.0\times10^{-23}$~~~~&~~$\mu\tau\footnotemark[6]$~~~~&$0.5\times10^{-23}$ \\
${\mathcal Im}(c_{L})^{TY}_{ab}$~~~&~~$e\mu$~~~~&$9.0\times10^{-23}$~~~~&~~$\mu\tau$~~~~&$9.0\times10^{-23}$ \\ \hline
${\mathcal Re}(c_{L})^{XX}_{ab}$~~~&~~$e\mu$~~~~&$4.6\times10^{-21}$~~~~&~~$\mu\tau\footnotemark[6]$~~~~&$2.5\times10^{-23}$ \\
${\mathcal Im}(c_{L})^{XX}_{ab}$~~~&~~$e\mu$~~~~&$4.6\times10^{-21}$~~~~&~~$\mu\tau$~~~~&$4.6\times10^{-21}$ \\ \hline
${\mathcal Re}(c_{L})^{YY}_{ab}$~~~&~~$e\mu$~~~~&$4.5\times10^{-21}$~~~~&~~$\mu\tau\footnotemark[6]$~~~~&$2.4\times10^{-23}$ \\
${\mathcal Im}(c_{L})^{YY}_{ab}$~~~&~~$e\mu$~~~~&$4.5\times10^{-21}$~~~~&~~$\mu\tau$~~~~&$4.5\times10^{-21}$ \\ \hline
${\mathcal Re}(c_{L})^{XZ}_{ab}$~~~&~~$e\mu$~~~~&$1.1\times10^{-21}$~~~~&~~$\mu\tau\footnotemark[6]$~~~~&$0.7\times10^{-23}$ \\
${\mathcal Im}(c_{L})^{XZ}_{ab}$~~~&~~$e\mu$~~~~&$1.1\times10^{-21}$~~~~&~~$\mu\tau$~~~~&$1.1\times10^{-21}$ \\ \hline
${\mathcal Re}(c_{L})^{YZ}_{ab}$~~~&~~$e\mu$~~~~&$1.1\times10^{-21}$~~~~&~~$\mu\tau\footnotemark[6]$~~~~&$0.7\times10^{-23}$ \\
${\mathcal Im}(c_{L})^{YZ}_{ab}$~~~&~~$e\mu$~~~~&$1.1\times10^{-21}$~~~~&~~$\mu\tau$~~~~&$1.1\times10^{-21}$ \\ \hline
${\mathcal Re}(c_{L})^{XY}_{ab}$~~~&~~$e\mu$~~~~&$2.2\times10^{-21}$~~~~&~~$\mu\tau\footnotemark[6]$~~~~&$1.2\times10^{-23}$ \\
${\mathcal Im}(c_{L})^{XY}_{ab}$~~~&~~$e\mu$~~~~&$2.2\times10^{-21}$~~~~&~~$\mu\tau$~~~~&$2.2\times10^{-21}$ \\ \hline

\hline \hline

\end{tabular}
\footnotetext[6]{Determined using FD data~\cite{paper2}.}
\end{table}
As discussed, the way we determine the upper limits does not distinguish between the real and imaginary parts of the SME coefficients for the oscillation processes $\nu_\mu \rightarrow \nu_e$ and $\nu_\mu \rightarrow \nu_\tau$.  This is reflected in Table~\ref{table:limits2}.

We compare the 36 limits in Table~\ref{table:limits2} with those determined by LSND and IceCube.  In \cite{ndlv}, we showed that the MINOS upper limits determined with only ND neutrino data were already more sensitive than those found by LSND~\cite{LSND}.  IceCube analyzed their data using the simple ``vector model"~\cite{KM2} for the real components of four SME coefficients for $\nu_{\mu} \rightarrow \nu_\tau$ transitions, giving ${\mathcal Re}(a_{L})^{X}_{\mu\tau}$, ${\mathcal Re}(a_{L})^{Y}_{\mu\tau}$ $< 1.8 \times 10^{-23}$ GeV and ${\mathcal Re}(c_{L})^{TX}_{\mu\tau}$, ${\mathcal Re}(c_{L})^{TY}_{\mu\tau}$ $< 3.7 \times 10^{-27}$ \cite{Abbasi:2010kx}.  The IceCube $a_{L}$-type limits are a factor of 3 lower and the $c_{L}$-type limits 4 orders of magnitude lower than the MINOS limits reported here for these four coefficients.


We have presented a search for the Lorentz and CPT violating sidereal signal predicted by the SME theory with antineutrinos detected in the MINOS Near Detector.   We found no significant evidence for sidereal variations in a blind analysis of the data.  Furthermore, the effects of systematic uncertainties on these results are not significant.  When framed in the SME theory~\cite{KM}, these results lead to the conclusion that we have detected no evidence for Lorentz invariance violation in the antineutrino data set.  While the large number of coefficients describing the theory could fortuitously cancel a sidereal signal, the MINOS antineutrino and neutrino results, when taken together, suggest that this is improbable.  

We computed upper limits for the 36 SME coefficients appropriate to this analysis.  We then combined these with the upper limits we found in our previous analyses, and the results are given in Table~\ref{table:limits2}.  MINOS provides the lowest limits for 32 of these coefficients.  

We gratefully acknowledge our many valuable conversations with Alan Kosteleck\'y and Jorge D\'iaz during the course of this work.  This work was supported by the US DOE, the UK STFC, the US NSF, the State and University of Minnesota, the University of Athens, Greece, and Brazil's FAPESP, CNPq and CAPES.  We are grateful to the Minnesota Department of Natural Resources, the crew of the Soudan Underground Laboratory, and the staff of Fermilab for their contributions to this effort.

\bibliography{lorentzAntiND_paper}

\end{document}